\documentclass[aps,pra,showpacs,floatfix,reprint]{revtex4-1}
\usepackage{bm}
\usepackage{graphicx} 
\usepackage{amsmath}
\usepackage{amsfonts}
\usepackage{color}
\begin{document}
\title{Multichannel Quantum Defect Theory for cold molecular collisions \\
with a strongly anisotropic potential energy surface}

\author{James F. E. Croft}
\author{Jeremy M. Hutson}
\affiliation{Joint Quantum Centre (JQC) Durham/Newcastle, Department of
Chemistry, Durham University, South Road, Durham, DH1~3LE, United Kingdom}

\date{\today}

\begin{abstract}
We show that multichannel quantum defect theory (MQDT) can be applied
successfully as an efficient computational method for cold molecular collisions
in Li+NH, which has a deep and strongly anisotropic interaction potential. In
this strongly coupled system, closed-channel poles restrict the range over
which the MQDT $\bm Y$ can be interpolated. We present an improved procedure to
transform the MQDT reference functions so that the poles are removed from the
energy range of interest. Effects due to very long-range spin-dipolar couplings
are outside the scope of MQDT, but can be added perturbatively. The new
procedure makes it possible to calculate the elastic and inelastic cross
sections needed to evaluate the feasibility of sympathetic cooling of NH by Li
using coupled-channel calculations at only 5 combinations of energy and field.
\end{abstract}

\maketitle

\section{Introduction}
\label{intro}

Cold and ultracold molecules provide an exciting doorway to new fields in
physics and chemistry. They share the controllability and tunable interactions
that have made cold atom studies so fruitful. However, the richer structure of
molecules makes them suitable for many new applications and research directions
\cite{Carr:NJPintro:2009, Quemener:2012}. Polar molecules are of particular
interest because their electric dipole moment allows the interactions between
them to be controlled and manipulated by external fields. This strong tunable
response may make it possible to develop a fully controlled chemistry
\cite{Krems:PCCP:2008} where every degree of freedom of the reaction can be
tuned, providing fundamental insights into chemical reaction processes.
Molecules such as KRb \cite{Ni:KRb:2008} and Cs$_2$ \cite{Danzl:ground:2010}
have already been produced at submicrokelvin temperatures in their
lowest-energy electronic, vibrational, rotational, and hyperfine state, by
magnetoassociation followed by laser state transfer. Reactions of ultracold
$^{40}$K$^{87}$Rb with itself and with K and Rb atoms have been studied
\cite{Ospelkaus:react:2010}, and it was seen that quantum statistics and
quantum threshold laws play an important role in determining the rates of
inelastic collisions.

The only ultracold molecules that are accessible with current methods are
alkali-metal dimers. However, there is great interest in producing samples of
other molecular species in the ultracold regime. The most promising route to
this is first to cool and trap them in the cold regime (at temperatures below
1~K), using a method such as buffer-gas cooling \cite{Doyle:2004} or molecular
beam deceleration \cite{Bethlem:IRPC:2003} and then to bring them to the
ultracold regime using a second-stage approach such as evaporative cooling
\cite{Stuhl:2012}, sympathetic cooling \cite{Soldan:2004}, or laser cooling
\cite{Shuman:2010, Hummon:2012}.

Evaporative and sympathetic cooling both rely on elastic collisions to
thermalize the sample, but both can be prevented by inelastic collisions that
release kinetic energy and lead to trap loss. Collisional properties are also
key to methods for {\em controlling} ultracold atoms and molecules.
Calculations on atomic and molecular collisions are therefore crucial to both
the production and control of cold and ultracold molecules. Such calculations
require the solution of the set of coupled differential equations obtained from
the Schr\"odinger equation. There are are various numerical methods for solving
these coupled differential equations, of which the most commonly used is the
full coupled-channels method. This propagates the matrix solution of the
Schr\"odinger equation from short to long range and takes a time proportional
to $N^3$, where $N$ is the number of coupled channels. The properties of
completed collisions are described by the scattering matrix $\bm S$, which is
obtained by matching the propagated solutions to free-particle wavefunctions
(Ricatti-Bessel functions) at long range \cite{Johnson:1973}.

Full coupled-channels calculations can be extremely expensive, particularly in
applied electric and magnetic fields \cite{Volpi:2002, Krems:mfield:2004,
Gonzalez-Martinez:2007}. The expense is particularly great for systems with
deep and strongly anisotropic potential wells \cite{Wallis:LiNH:2011}, for
molecule-molecule collisions \cite{Janssen:NHNH:field:2011}, or when nuclear
hyperfine interactions are included \cite{Lara:PRA:2007,
Gonzalez-Martinez:hyperfine:2011}.

In cold collision studies, the scattering $\bm S$ matrix is often a fast
function of collision energy $E$ and magnetic field $B$, with extensive
structure due to scattering resonances and discontinuous behavior at threshold.
Calculations are thus required over a fine grid of energies and/or applied
field, typically over an energy range of order 1~K, from threshold, and for
magnetic fields up to a few thousand G. This contrasts with the situation for
collisions of ultracold atoms, where the energy range of interest is commonly a
few $\mu$K and the fields are typically a few hundred G.

Approaches based on multichannel quantum defect theory (MQDT) avoid the
repetition of the expensive propagation by representing the scattering
properties in terms of a matrix $\bm Y(E,B)$ that is a smooth function of $E$
and $B$ \cite{Greene:1979, Greene:1982, Mies:1984a, Mies:MQDT:2000,
Raoult:2004}. MQDT has proved highly successful for cold atomic interactions
\cite{Gao:2008, Julienne:FD142, Gao:QDT:1998, Julienne:1989, Burke:1998,
Mies:MQDT:2000, Raoult:2004}, and more recently it has been applied to
collisions of cold and ultracold molecules \cite{Idziaszek:PRL:2010,
Gao:react:2010, Idziaszek:PRA:2010, Croft:MQDT:2011, Mayle:2012,
Croft:MQDT2:2012}. MQDT defines the matrix $\bm{Y}(E,B)$ at a matching
distance, $r_{\rm match}$, at relatively short range in order to achieve this
insensitivity to energy and field. The $\bm Y$ matrix contains all the
scattering dynamics inside $r_{\rm match}$. The smooth variation of $\bm Y$
allows it to be obtained once and then used for calculations over a wide range
of energies and fields, or obtained by interpolation from a few points. The
computational cost of calculations at additional energies and fields is only
proportional to $N$, not $N^3$.

We have previously demonstrated the application of MQDT to cold molecular
collisions for the moderately anisotropic system Mg+NH($^3\Sigma^-$)
\cite{Croft:MQDT:2011, Croft:MQDT2:2012}. In ref.\ \cite{Croft:MQDT2:2012}, we
showed that the choice made for the {\em phase} of the MQDT reference functions
is very important in producing a $\bm Y$ matrix that can be interpolated
smoothly over a wide range of energy and field. The purpose of the present
paper is to explore how the approach performs for a much more strongly
anisotropic system with many more closed channels. For this purpose we choose
Li+NH, which has been studied previously using full coupled-channels
calculations by Wallis {\em et al.}\ \cite{Wallis:LiNH:2011}. As in ref.\
\cite{Wallis:LiNH:2011}, we focus on collisions between spin-polarized Li and
NH, which occur on the quartet potential energy surface. This surface is deep
and highly anisotropic, with a well depth about 1800 cm$^{-1}$ at the Li-NH
geometry but only 113 cm$^{-1}$ at the NH-Li geometry. With a small but
important modification to the method of choosing phases described in ref.\
\cite{Croft:MQDT2:2012}, we find that we can obtain accurate results for
elastic and inelastic cross section, over the entire range of energies relevant
to sympathetic cooling, using only 5 coupled-channels calculations.

\section{Theory}
The theory of MQDT is given in details in refs.\ \cite{Greene:1982, Mies:1984a,
Mies:MQDT:2000, Raoult:2004}. Here we give only a brief description, following
references \cite{Croft:MQDT:2011, Croft:MQDT2:2012}, which is sufficient to
describe the notation we use.

MQDT makes the approximation that the multichannel Schr\"odinger equation is
uncoupled at distances $r>r_{\rm match}$, so that its solution in this region
may be written in the matrix form
\begin{equation}\label{eqn:match_mqdt}
 \bm{\Psi} = r^{-1} \left[\bm{f}(r) + \bm{g}(r)\bm{Y}\right].
\end{equation}
Here $\bm{f}$ and $\bm{g}$ are diagonal matrices containing the functions $f_i$
and $g_i$, which are linearly independent solutions of a reference
Schr\"odinger equation in each asymptotic channel $i$,
\begin{equation}
\left[-\frac{\hbar^2}{2\mu}\frac{d^2}{dr^2} + U_i^{\rm ref}(r) - E\right] f_i(r) =0,
\label{eq:SEref}
\end{equation}
and similarly for $g_i(r)$. $\mu$ is the reduced mass, and the reference
potentials $U_i^{\rm ref}(r)$ approach the true potential at long range. They
include the centrifugal terms $\hbar^2 L_i(L_i+1)/2\mu r^2$, where $L_i$ is the
partial-wave quantum number for channel $i$.

The phase of the short-range reference functions $f_i$ and $g_i$ is a
disposable parameter of MQDT and may be chosen to generate a $\bm Y$ matrix
that is smooth and pole-free over a wide range of energy and field
\cite{Croft:MQDT2:2012}. Equation (\ref{eqn:match_mqdt}) shows that the $\bm Y$
matrix has a pole whenever the component of the propagated multichannel
wavefunction $\psi_i$ in any channel $i$ is proportional to the reference
function $g_i$ and has no contribution from $f_i$, i.e.\ when $g_i$ and the
full coupled-channels solution have the same phase at $r_{\rm match}$. In ref.\
\cite{Croft:MQDT2:2012}, we proposed transforming the pair of reference
functions $f_i$ and $g_i$ in channel $i$ with a rotation angle $\theta_i$,
chosen so that the diagonal matrix elements $Y_{ii}$ is 0 at a particular
reference energy $E_{\rm ref}$ and field $B_{\rm ref}$. This ensures that the
reference function $g_i$ and the full coupled-channels solution in channel $i$
are perfectly out of phase at the chosen $r_{\rm match}$, and the resulting
$\bm Y$ matrix is therefore pole-free close to $E_{\rm ref}$ and $B_{\rm ref}$
\cite{Croft:MQDT2:2012}.

The range of the pole-free region is dependent on where the matching
occurs. When matching is in the classically allowed region, the phases
of both the reference functions and the propagated coupled-channels
solutions vary approximately linearly with energy and setting the
diagonal $\bm Y$ matrix elements to zero is effective: the {\em
relative} phase of the reference functions and the coupled-channels
solution is a slow function of energy. For a closed channel where
matching is carried out in the classically forbidden region, however,
there is resonance structure in both the coupled-channels solutions and
the reference functions so that the phase of each is a fast (and
nonlinear) function of energy. Fortunately, the energies at which the
resonance structure occurs depend on the choice of $\theta_i$. In the
present work we show that a more sophisticated choice of $\theta_i$
than that of ref.\ \cite{Croft:MQDT2:2012} can produce a larger
pole-free region for closed channels.

\subsection{Basis set and quantum numbers}

We construct the collision Hamiltonian in the fully uncoupled basis set
$|nm_n\rangle |s_{\rm NH} m_{s_{\rm NH}}\rangle|s_{\rm Li} m_{s_{\rm
Li}}\rangle|LM_L\rangle$, where the quantum numbers $n$ and $s_{\rm NH}$
describe the rotation and electron spin of the NH molecule and $s_{\rm Li}$
describes the electron spin of the Li atom. The corresponding $m$ quantum
numbers are the projections onto the space-fixed magnetic field axis. Hyperfine
structure is neglected. The matrix elements required for the coupled equations
are the same as for scattering of NH from a closed-shell atom
\cite{Gonzalez-Martinez:2007}, with the addition of the anisotropic
intermolecular spin-spin dipolar interaction \cite{Wallis:LiNH:2011}.

The coupled-channels calculations may in principle be carried out in any
sufficiently complete basis set. However, the $\bm Y$ and $\bm S$ matrices are
expressed in a basis set of eigenfunctions of the field-dressed monomer
Hamiltonians. At low field the states of the NH molecule are approximately
described by quantum numbers $j$ and $m_j$, where $j$ is the resultant of $n$
and $s$. We label elements of $\bm Y$ and $\bm S$ by subscripts $n, j, m_j,
m_{s_{\rm Li}}, L,M_L \rightarrow n', j', m'_j, m'_{s_{\rm Li}}, L',M'_L $. For
diagonal elements we suppress the second set of labels.

\subsection{Numerical methods}

The coupled-channels calculations required for both MQDT and the full
coupled-channels approach are carried out using the MOLSCAT package
\cite{molscat:v14}, as modified to handle collisions in magnetic fields
\cite{Gonzalez-Martinez:2007}. The coupled equations are solved numerically
using the hybrid log-derivative propagator of Alexander and Manolopoulos
\cite{Alexander:1987}, which uses a fixed-step-size log-derivative propagator
in the short-range region ($r_\text{min} \le r < r_\text{mid}$) and a
variable-step-size Airy propagator in the long-range region ($r_\text{mid} \le
r \le r_\text{max}$). As in ref.\ \cite{Wallis:LiNH:2011}, the full
coupled-channels calculations use $r_\text{min}=1.8$~\AA,
$r_\text{mid}=12.5$~\AA\ and $r_\text{max}=600$~\AA\ (where 1~\AA\ =
$10^{-10}$~m). MQDT requires coupled-channels calculations only from
$r_\text{min}$ to $r_{\rm match}$ (which is less than $r_\text{mid}$), so only
the fixed-step-size propagator is used in this case.

The initial MQDT reference functions and quantum defect parameters are
obtained as described in ref.\ \cite{Croft:MQDT:2011}, using the
renormalized Numerov method \cite{Johnson:1977} to solve the
1-dimension Schr\"odinger equations for the reference potentials. In
this paper all MQDT calculations use the reference potential
\begin{equation}\label{eqn:refunc3} U^{\text{ref}}_i(r) = V_0(r) +
\frac{\hbar^2 L_i(L_i+1)}{2\mu r^2} + E_i^\infty,
\end{equation}
where $V_0(r)$ is the isotropic part of the interaction potential and $L_i$ is
the partial-wave quantum number for channel $i$. The reference potential
contains a hard wall at $r = r^{\text{wall}}_i$, so that $U^\text{ref}_i(r) =
\infty$ for $r<r^{\text{wall}}_i$. In the present paper we choose
$r^{\text{wall}}_i = 4.0$~\AA. This reference potential has been shown to
produce quantitatively accurate results for Mg+NH \cite{Croft:MQDT:2011,
Croft:MQDT2:2012}.

The uncoupled basis functions used to solve the coupled-channel equations are
not eigenfunctions of the Hamiltonians of the separated monomers. The
log-derivative matrix obtained from the coupled-channel calculations at a
distance $r_\text{match}$ is therefore transformed into a basis set that
diagonalizes the asymptotic Hamiltonian \footnote{This transformation was
neglected in refs.\ [34] and [36],
where the partially coupled basis set used for the propagation was
asymptotically very nearly diagonal. However, it is important in the decoupled
basis set used here.}. The MQDT $\bm{Y}$ matrix is then obtained by matching to
this log-derivative matrix at $r_\text{match}$ using Eq.\ (13) of ref.\
\cite{Croft:MQDT:2011}. All channels with $n \ge 2$ are treated as strongly
closed and thus not included in the MQDT part of the calculation, but are
included in the log-derivative propagation.

\section{Results and discussion}

In a magnetic field, the lowest Li-NH threshold ($n=0,s_{\rm
NH}=1,s_{\rm Li}=\frac{1}{2}$) splits into 6 Zeeman sublevels. We
consider collisions between Li atoms and NH molecules that are both
initially in their magnetically trappable low-spin-seeking states,
$m_{s,{\rm Li}}=+1/2$ and $m_{s,{\rm NH}}=+1$. This corresponds to the
highest of the 6 thresholds.

\begin{figure}[tb]
\centering
\includegraphics[width=1\columnwidth]{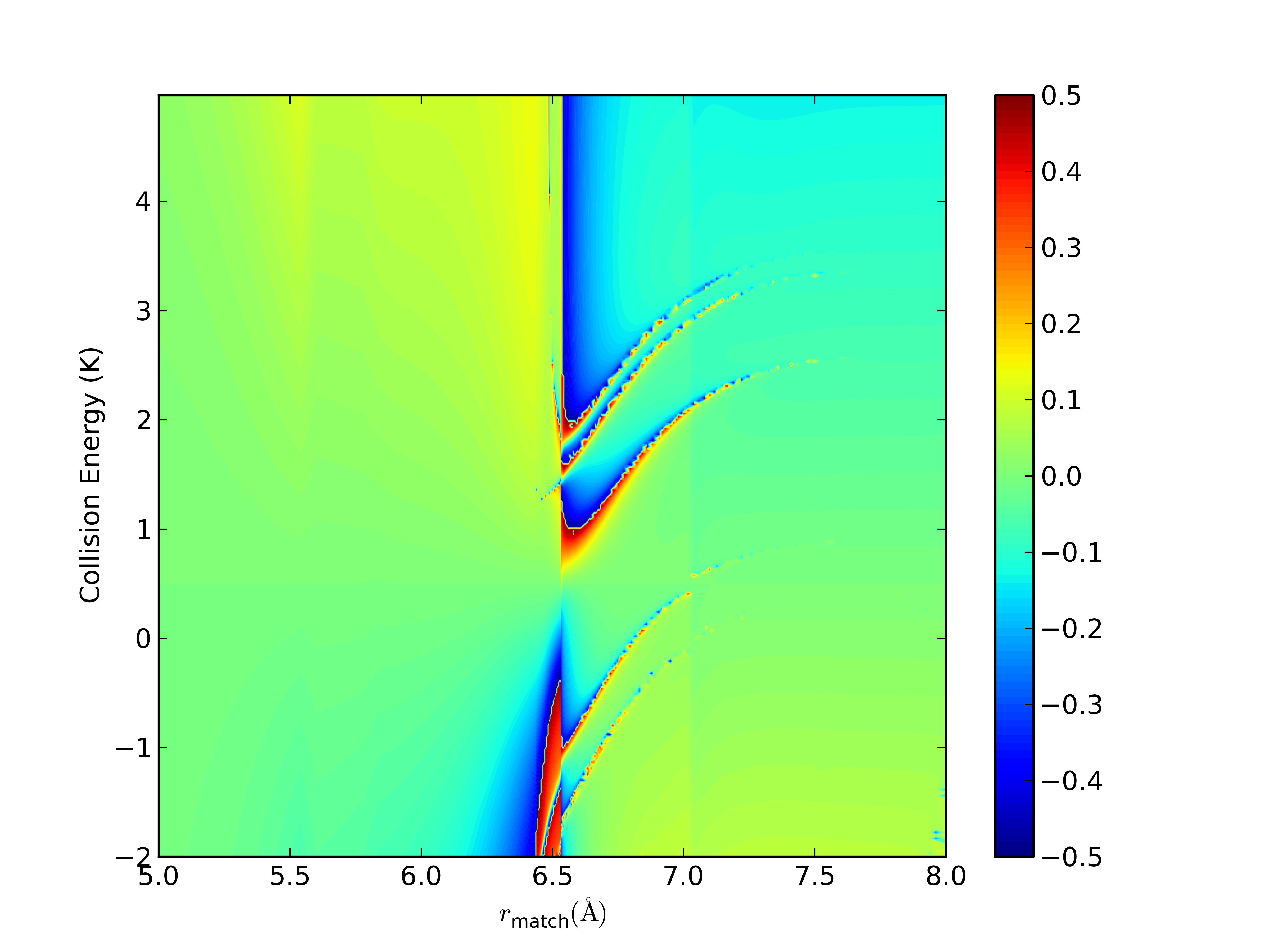}
\caption{(Color online) $\arctan Y_{ii} / \pi$ for a single representative
diagonal $\bm Y$ matrix element, as a function of collision energy $E$ and
$r_{\rm{match}}$, with the phases $\theta_i$ set so that $Y_{ii}=0$ in all
channels for energy $E_{\rm ref}=0.5$~K and field $B_{\rm ref}=10$~G.}
\label{fig:Y_OPT_ALLR}
\end{figure}

Figure \ref{fig:Y_OPT_ALLR} shows the variation of the representative element
$Y_{1, 2, -2, -\frac{1}{2}, 7, 4}$, as a function of the matching distance and
energy, when the phases $\theta_i$ are chosen to make all diagonal $\bm Y$
matrix elements zero at collision energy $E_{\rm ref}=0.5$~K and field $B_{\rm
ref}=10$~G. It may be seen that there are poles (where $\arctan Y_{ii} / \pi$
passes through $\pm1/2$) whose positions depend strongly on $r_{\rm match}$.
Other $\bm Y$ matrix elements are quantitatively different but have poles in
the same places. The basis set used for Figure \ref{fig:Y_OPT_ALLR} includes
all functions up to $n_{\rm max} = 3$ and $L_{\rm max} = 8$. This unconverged
basis set was used due to the substantial computational cost of performing a
full coupled-channels calculation at every energy in the Figure. The outer
turning point of the $n=1$ reference potential is at $6.1$~\AA, and it may be
seen that, for values of $r_{\rm match}$ inside this, the $\bm Y$ matrix is
pole-free over many K. However, MQDT with such small values of $r_{\rm match}$
does not produce accurate results because it neglects all channel couplings
that exist outside $r_{\rm match}$. When $r_{\rm match} > 6.1$~\AA, poles start
to enter the $\bm Y$ matrix in the energy range of interest. As $r_{\rm match}$
increases further, the poles move and at some values of $r_{\rm match}$ can
come within 0.1~K of $E_{\rm ref}$.

A contour plot such as Fig.\ \ref{fig:Y_OPT_ALLR} requires coupled-channels
calculations at every energy, and producing it thus sacrifices most of the
computational savings that MQDT is designed to achieve. In addition, we need a
procedure for choosing the phases $\theta_i$ that will guarantee a large
pole-free region for {\em any} choice of $r_{\rm match}$. In the remainder of
this paper, we perform calculations at only a single value of $r_{\rm
match}=6.5$~\AA, deliberately chosen to be in a region where Fig.\
\ref{fig:Y_OPT_ALLR} shows that there are poles in $\bm Y$ inconveniently close
to $E_{\rm ref}$. In addition, the remaining calculations use a basis set
including all functions up to $n_{\rm max} = 6$ and $L_{\rm max} = 8$, except
where stated otherwise.

\begin{figure}[tb]
\centering
\includegraphics[width=1\columnwidth]{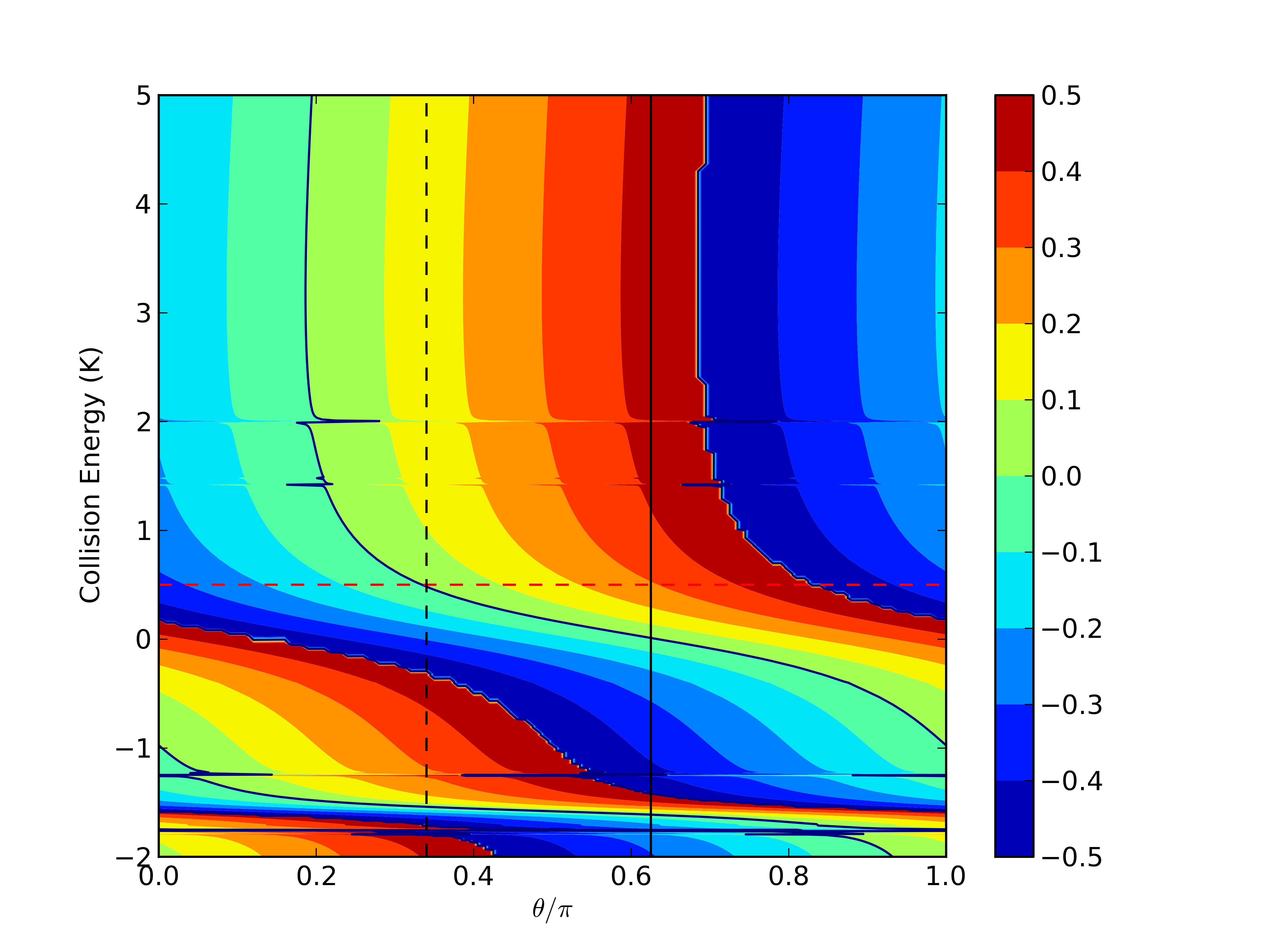}
\includegraphics[width=1\columnwidth]{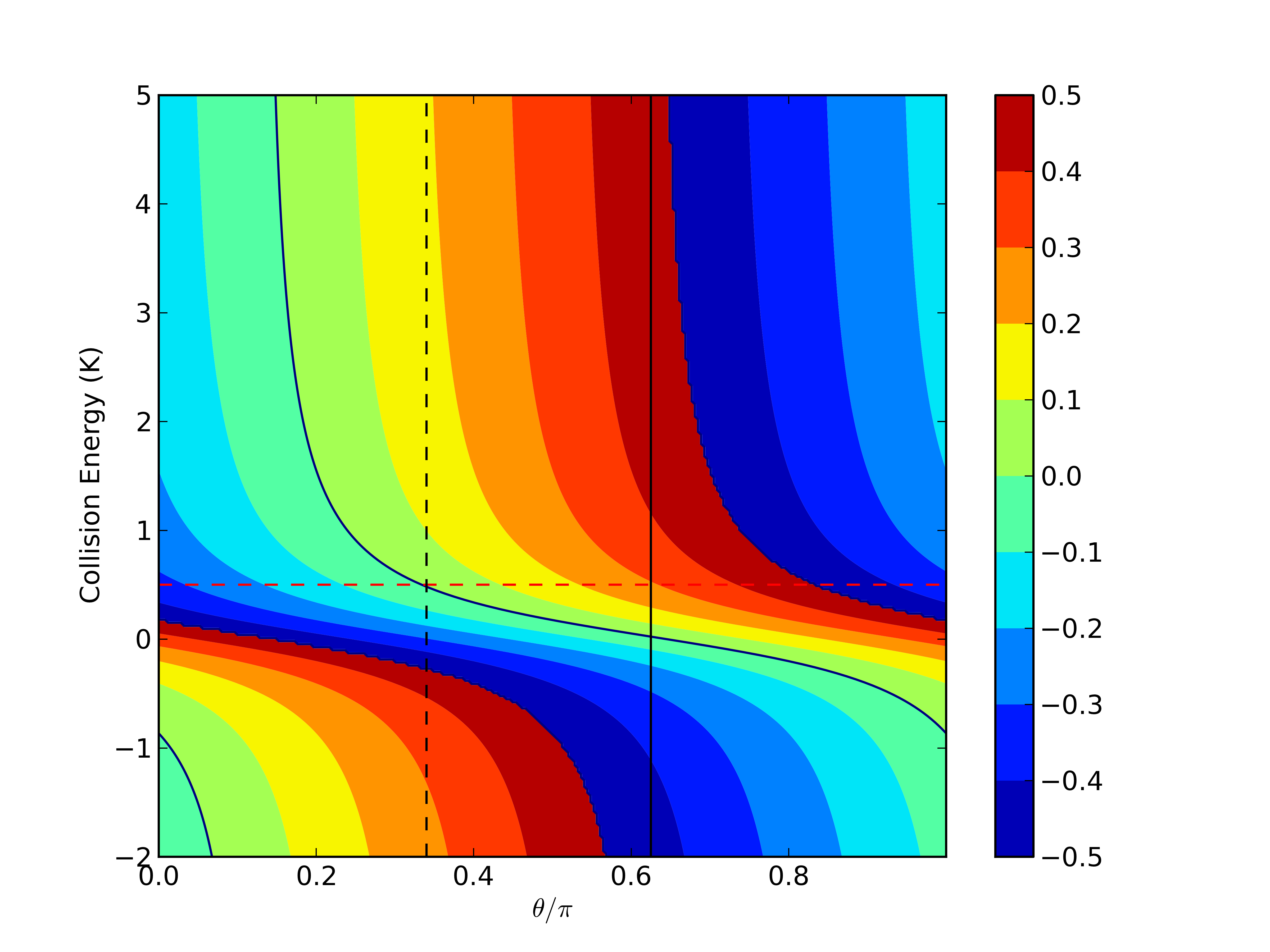}
\caption{(Color online) Upper panel: $\arctan Y_{ii} / \pi$ from
coupled-channels calculations at $B=10$~G for a single diagonal element, $Y_{1,
2, -2, -\frac{1}{2}, 7, 4}$, with $\theta_{i'}$ set so that $Y_{i'i'}=0$ in all
other channels $i'$ at $E_{\rm ref}=0.5$~K (horizonal dashed line). Lower panel:
$\arctan Y_{ii} / \pi$ as a function of energy and $\theta_i$ for a single
uncoupled channel as given by Eqs.\ (\ref{eqn:tan}) and (\ref{eqn:phase}).
} \label{fig:Theta_scan}
\end{figure}

The dependence of a diagonal $\bm Y$ matrix element on the phase of the
reference functions (in any channel handled by MQDT) is
\begin{equation}
\label{eqn:tan}
Y_{ii}(E) = \tan\left[\theta_i + \delta_i(E)\right],
\end{equation}
where $\delta_i(E)$ is the phase shift between the unrotated reference
function $f_i$ and the propagated multichannel wavefunction in channel
$i$. For a closed channel that is capable of supporting resonances, the
phase shift around a resonance has a Breit-Wigner form,
\begin{equation}
\label{eqn:phase}
\delta_i(E) = \bar{\delta_i}(E) + \arctan \left(\frac{\Gamma_i/2}{E-E_i^{\rm res}}\right),
\end{equation}
where $\bar{\delta_i}$ is a slowly varying (non-resonant) background term. The
resonant part of this function is shown in the lower panel of Fig.\
\ref{fig:Theta_scan}, for values of the parameters appropriate to one of the
channels in Li+NH. It may be seen that choosing a value of $\theta_i$ that
makes $Y_{ii}$ zero (shown by the dashed vertical line) does not guarantee a
large pole-free region in the case where $E_{\rm ref}$ is close to $E_i^{\rm
res}$. A much better choice in this case is to set $\theta_i$ to the value
shown by the solid vertical line. In the following we will show how this can be
achieved.

A basic problem of MQDT in coupled-channel problems is that a pole in $\bm Y$
that originates in {\em any} channel causes a pole in {\em every} channel. We
refer to this as the {\em contamination} of one channel by another. The upper
panel of Figure \ref{fig:Theta_scan} shows $\arctan Y_{ii} / \pi$ for the
single matrix element, $Y_{1, 2, -2, -\frac{1}{2}, 7, 4}$, obtained from
coupled-channels calculations as a function of $\theta_i$ and collision energy
$E$. The phases $\theta_{i'}$ in all other channels $i'$ are set to the values
that produce $Y_{i'i'}=0$ at $E_{\rm ref}=0.5$~K. The broad horizontal sweep
around 0~K arises from a resonance in channel $i$, while the narrower sweeps at
$-1.8$~K, $-1.6$~K, $-1.2$~K 1.4~K and 2.0~K are poles due to contamination
from other channels. Setting $Y_{i'i'}$ to zero in all these other channels has
shifted these contamination effects to energies either above about 1.4~K or
below $-1.2$~K , leaving a region of about 2.6~K uncontaminated by other
channels. For the specific circumstances shown in Figure \ref{fig:Theta_scan},
it is seen that choosing $\theta_i$ so that $Y_{ii} = 0$ results in
$\theta_i/\pi\approx0.33$ (dashed vertical line). This produces a pole in
$Y_{ii}$ itself at about $-0.1$~K, which is inconveniently close to $E_{\rm
ref}$. However, a choice of $\theta_i/\pi=0.63$ (solid vertical line) would
produce a much larger pole-free range, limited only by poles due to
contamination effects in other channels. If improved values of $\theta_{i'}$
can also be obtained for these contamination poles, then there is clearly the
prospect of achieving a much improved pole-free region.

The pole structure in channel $i$ when uncontaminated by other channels is
given by Eqs.\ (\ref{eqn:tan}) and (\ref{eqn:phase}). In order to use these to
obtain a better choice of $\theta_i$, we need values for $E_i^{\rm res}$,
$\Gamma_i$ and $\bar\delta_i$. To obtain these we first optimize the phases as
in ref.\ \cite{Croft:MQDT2:2012}, transforming the reference functions so that
$Y_{ii} = 0$ in all channels at energy $E_{\rm ref}$. This provides at least a
small region where $Y_{ii}$ is uncontaminated by poles in other channels. We
then carry out coupled-channels calculations at 2 additional energies near
$E_{\rm ref}$, and use Eqs.\ (\ref{eqn:tan}) and (\ref{eqn:phase}) to obtain
the three parameters $E_i^{\rm res}$, $\Gamma_i$ and $\bar\delta_i$
numerically, neglecting the slow variation of $\bar\delta_i$ with $E$. The
optimum pole-free region for this channel is then achieved by setting $\theta_i
= \pi/2-\bar\delta_i$.

The pole-free region for the entire $\bm Y$ matrix is optimized by applying
this procedure in all channels where there is resonant structure close to the
reference energy. We first calculate the numerical second derivative of the
diagonal matrix elements $Y_{ii}$ with respect to energy. We then select the
channel with the largest second derivative, apply the procedure described
above, and use the new set of phases to recalculate the three $\bm Y$ matrices.
Because of channel mixing, this in principle changes {\em all} the diagonal
matrix elements. If it reduces $\sum_i |d^2Y_{ii}/dE^2|$ then we accept the new
value of $\theta_i$. If not, we move on to the next channel and apply the same
procedure. We loop over the channels in this manner until there is no channel
for which changing $\theta_i$ to $\pi/2-\bar\delta_i$ reduces $\sum_i
|d^2Y_{ii}/dE^2|$. This is an inexpensive procedure, as it uses the same 3
coupled-channels calculations as before. Only the closed channels need to be
included in the loop since only these channels have resonance structure.

\begin{figure}[tb]
\centering
\includegraphics[width=1\columnwidth]{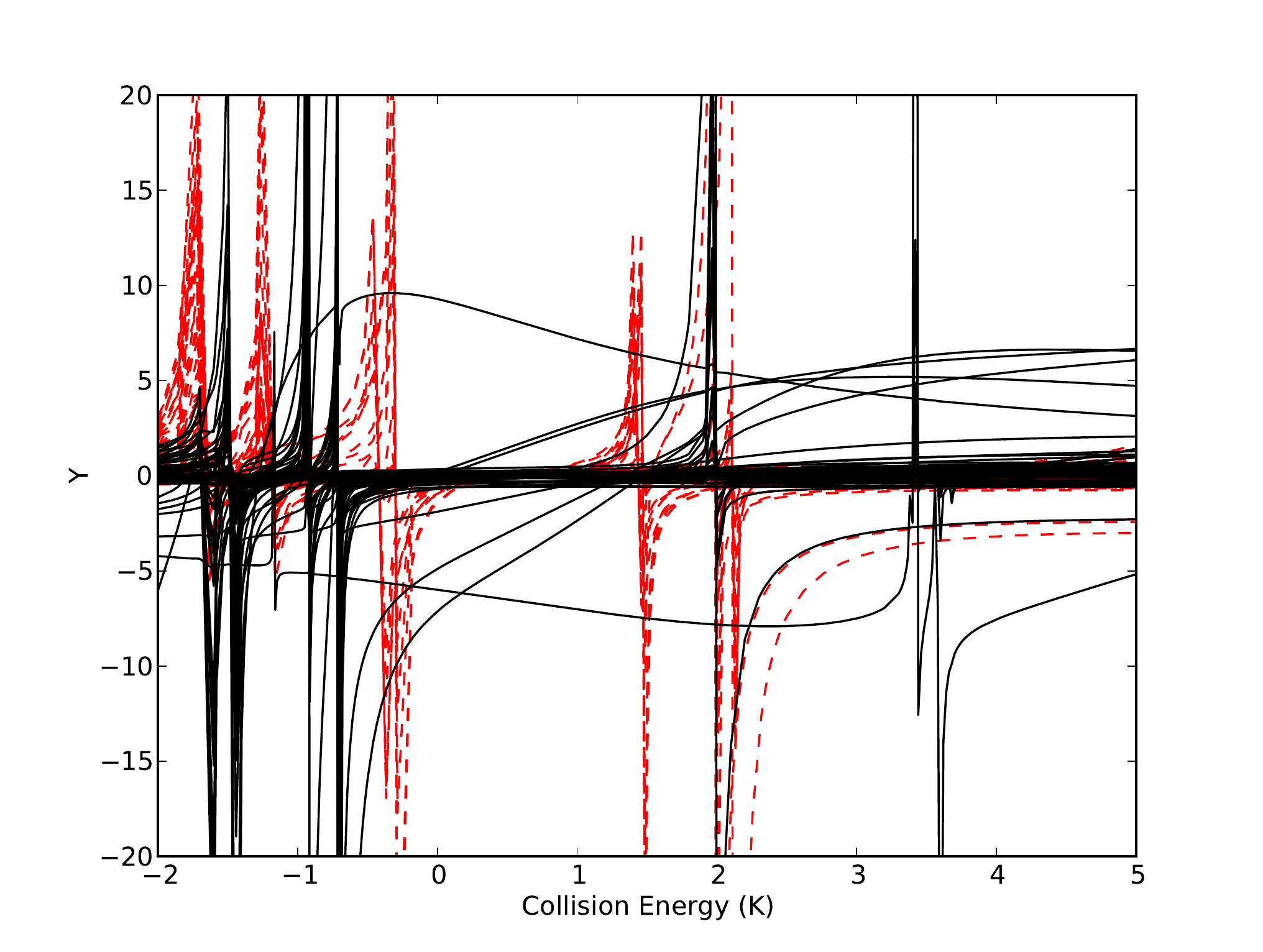}
\caption{(Color online) $Y_{ii}$ as a function of energy for all channels, with
all phases $\theta_i$ optimized as described in ref.\ \cite{Croft:MQDT:2011}
(dashed red lines) or using Eqs.\ (\ref{eqn:tan}) and (\ref{eqn:phase}) (solid
black lines). In both cases the phases are optimized at $r_{\rm match}=6.5$~\AA\
for $E_{\rm ref}=0.5$~K and $B_{\rm ref}=10$~G.} \label{fig:Y_DIAG}
\end{figure}
Figure \ref{fig:Y_DIAG} compares the final matrix elements $Y_{ii}$ in all the
channels included in the MQDT procedure, obtained with the two optimization
schemes. The dashed red lines show the result of choosing $\theta_i$ so that
$Y_{ii}$ is zero in every channel, as in ref.\ \cite{Croft:MQDT2:2012}, while
the solid black lines show the result of optimizing $\theta_i$ as described
above. Both calculations use $r_{\rm match}=6.5$~\AA\ and optimize $\theta_i$
at $E_{\rm ref}=0.5$~K and $B_{\rm ref}=10$~G. It may be seen that taking
account of closed-channel resonances significantly increases the pole-free
range of $\bm Y$. Furthermore, it produces $\bm Y$ matrix elements that are
considerably more linear between 0 and 1~K and may thus be interpolated more
accurately.

\begin{figure}[tb] \centering
\includegraphics[width=1\columnwidth]{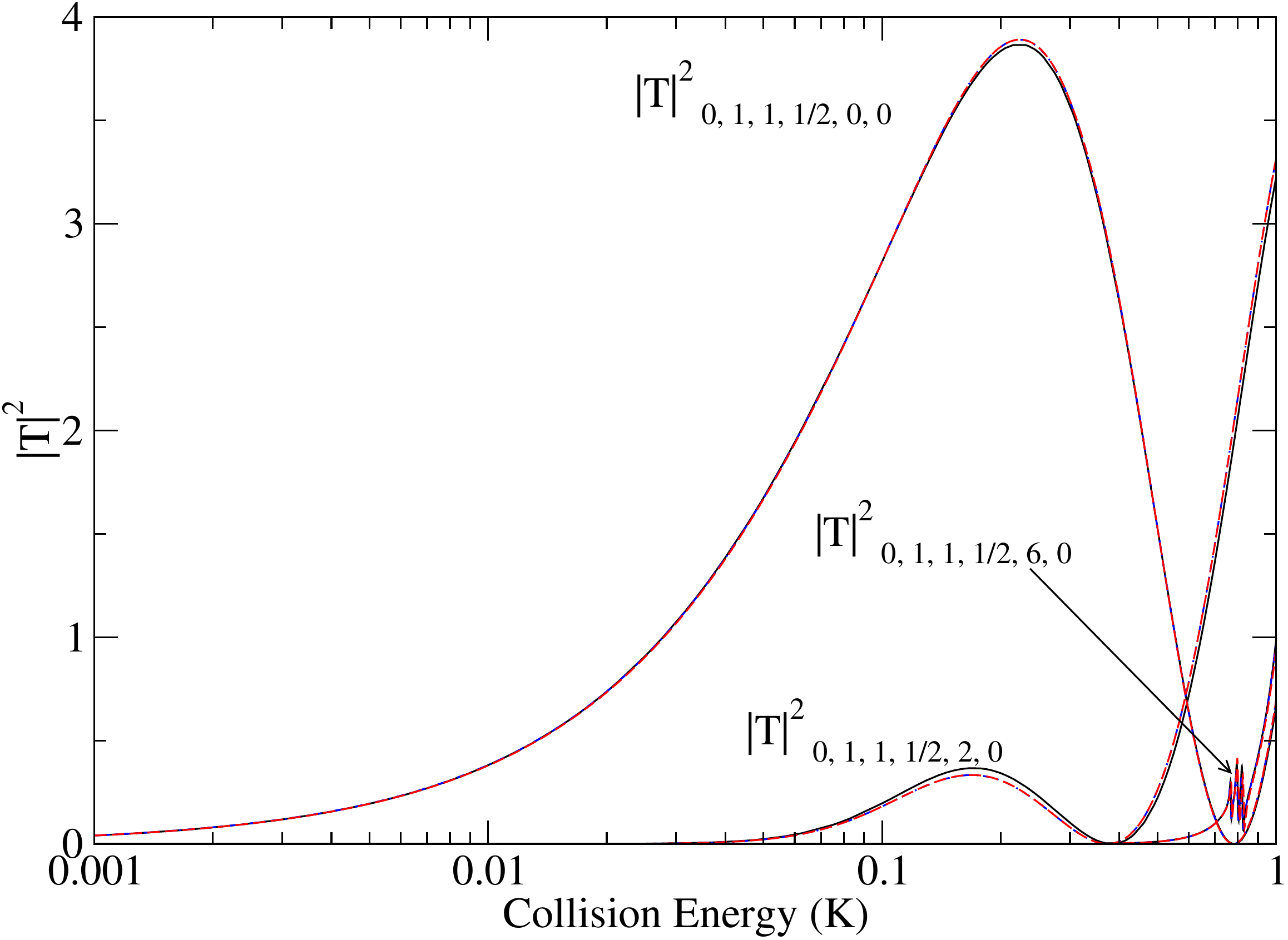}

\caption{(Color online) Squares of diagonal T-matrix elements $T_{{m_j},L,M_L}$
in the incoming channels for $m_j = +1$ and $L = 0$, 2 and 6 at $B = 10$ G,
obtained from full coupled-channels calculations (solid, black) and MQDT using
optimized reference functions for $r_{\rm{match}}$ = 6.5 \AA\ both with
(dot-dash, blue) and without (dashed, red) interpolation. $L=4$ is not shown
because it obscures the resonant feature for $L=6$. } \label{fig:S}
\end{figure}

Figure \ref{fig:S} compares diagonal T-matrix elements $|T_{ii}|^2$ (where
$T_{ij}=\delta_{ij}-S_{ij}$) obtained from full coupled-channels calculations
with those from the MQDT method using interpolation. The MQDT results were
obtained by interpolating (and extrapolating) $\bm Y$ quadratically using 3
points separated by 0.1~K around 0.5~K. The MQDT results obtained by
interpolation are very similar to the full coupled-channels results, even
around the resonance features at $E \approx 0.7$~K.

\section{Application to sympathetic cooling}
The key quantity that determines whether sympathetic cooling can
succeed is the ratio $\gamma$ of elastic to inelastic cross sections
for collisions of trapped molecules with coolant atoms. This ratio
typically needs to be greater than about 100 if trapped molecules are
to undergo enough elastic (thermalizing) collisions to achieve cooling
before undergoing an inelastic collision that releases kinetic energy
and causes trap loss. Wallis {\em et al.} \cite{Wallis:LiNH:2011} have
investigated Li + NH collisions using coupled-channel calculations, and
produced contour plots that show the ratio $\gamma$ as a function of
collision energy $E$ and magnetic field $B$. These calculations were
very expensive because they required calculations on a fine grid of
energies and fields. The contour plots given in ref.\
\cite{Wallis:LiNH:2011} actually included coupled-channel calculations
at 204 combinations of energy and field.

\begin{figure}[tb] \centering
\includegraphics[width=1\columnwidth]{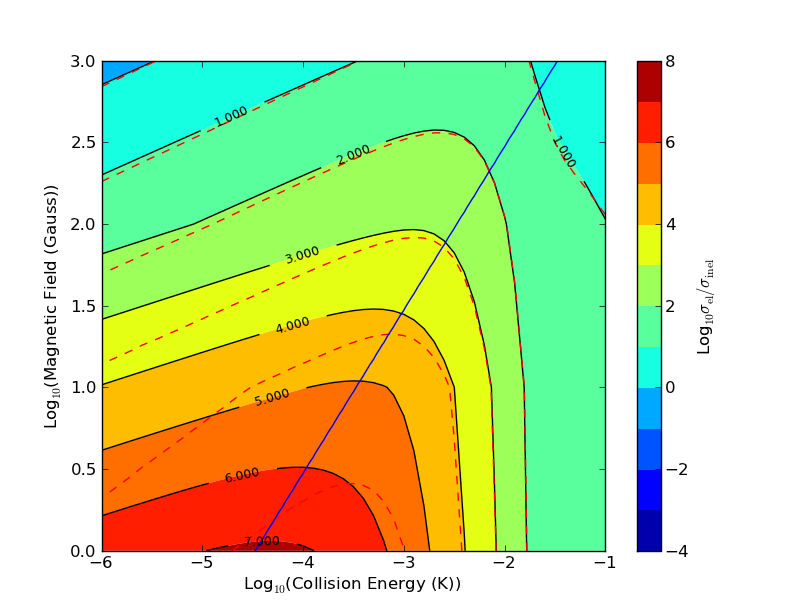}
\includegraphics[width=1\columnwidth]{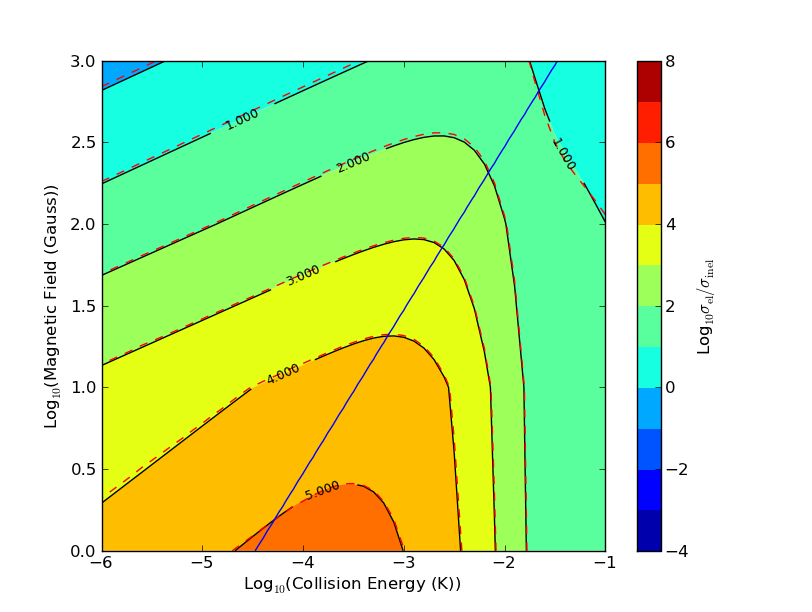}
\caption{(Color online) Contour plots of the ratio $\gamma$ of elastic to
inelastic cross sections for Li+NH collisions as a function of energy and
magnetic field, obtained using a basis set with $n_{\rm max}=6$ and $L_{\rm
max}=8$. The solid contours and shading show the MQDT results and the dashed
contours show the results from full coupled-channel calculations
\cite{Wallis:LiNH:2011}. The diagonal blue lines show the field at which the
Zeeman energy in a quadrupole trap is $6k_{\rm B}T$. Upper panel: results from
MQDT alone. Lower panel: results from MQDT with long-range spin-spin couplings
added using the hybrid Born approach.} \label{fig:sympa}
\end{figure}

MQDT offers the possibility of producing the entire contour plot from
coupled-channels calculations at just three energies (required to optimize the
phases) and two magnetic fields. We have therefore used MQDT to repeat the
calculations of ref.\ \cite{Wallis:LiNH:2011} on the unscaled potential energy
surface. Coupled-channel calculations from $r_{\rm min}$ to $r_{\rm match}$
were carried out at collision energies $E=0.01$, 0.05 and 0.1~K t magnetic
field $B=10$~G to optimize the phases, and at $E=0.01$ and 0.1~K at $B=1000$~G
to allow linear interpolation in $B$ and $E$. These calculations used a basis
set with $n_{\rm max}=6$ and $L_{\rm max}=8$ to allow direct comparison with
ref.\ \cite{Wallis:LiNH:2011}. The resulting contour plot of $\gamma$ is
compared with the results of ref.\ \cite{Wallis:LiNH:2011} in the upper panel
of Figure \ref{fig:sympa}: it may be seen that there is good agreement at both
high collision energies ($E>0.01$~K) and high fields ($B>100$~G), but that MQDT
by itself breaks down when both $E$ and $B$ are small \footnote{For consistency
of comparison, we have performed MQDT calculations on the same grid of energies
and fields as was used for the coupled-channels calculations in ref.\ [17]
This actually included only 4 field values across the range shown. Including
extra fields in the MQDT calculations produces small but visible changes in the
contours.}.

The inaccuracy in MQDT at low energy and low field occurs because, in this
region, the inelastic cross sections are dominated by {\em long-range}
inelasticity involving the magnetic dipole interaction between the spins of Li
and NH. As described by Janssen {\em et al.} \cite{Janssen:NHNH:field:2011},
there are long-range avoided crossings between the effective potential curves
for the incoming channel and for inelastic channels with larger values of $L$.
These crossings usually occur {\em outside} the centrifugal barriers, and even
coupled-channel calculations must be propagated to very long range (hundreds of
\AA) to capture their effects. They are thus outside the scope of MQDT, which
neglects couplings outside $r_{\rm match}$.

The long-range couplings may however be included perturbatively at very little
expense. Janssen {\em et al.} \cite{Janssen:NHNH:Born:2011} have developed both
a simple Born approximation and a distorted-wave Born approximation (DWBA) for
calculating the inelastic cross sections due to these long-range couplings. The
simple Born approximation is stable to evaluate, but can give results up to a
factor of 2 in error for cross sections for initial $L=0$ because it does not
take account of the phase shift due to short-range interactions. The DWBA, by
contrast, is quite accurate for initial $L=0$ but can be unstable when $L$ and
$L'$ are both non-zero. However, in the latter case, short-range effects are
unimportant. We have therefore used a hybrid Born approximation, made up of the
DWBA for initial $L=0$ and the simple Born approximation for initial $L>0$.
When we add the resulting inelastic cross sections to those from MQDT, we
obtain the contour plot for the ratio $\gamma$ shown in the lower panel of
Figure \ref{fig:sympa}. It may be seen that this gives essentially complete
agreement with the full coupled-channel results.

\begin{figure}[tb] \centering
\includegraphics[width=1\columnwidth]{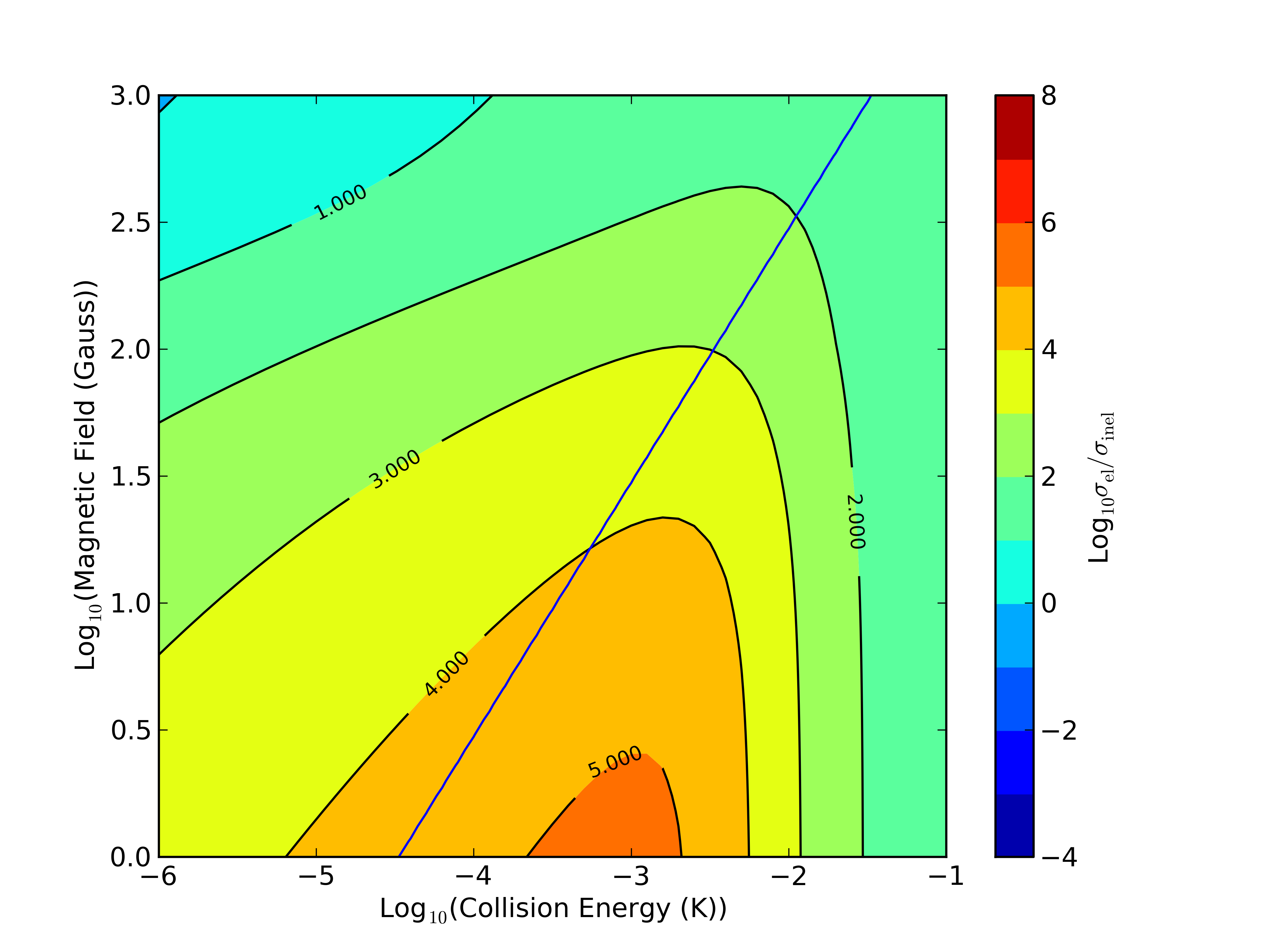}
\caption{(Color online) Contour plot of the ratio $\gamma$ of elastic to
inelastic cross sections for Li+NH collisions as a function of energy and
magnetic field, obtained from MQDT calculations with perturbative long-range
corrections, using a basis set with $n_{\rm max}=10$ and $L_{\rm max}=8$. The
diagonal blue line shows the field at which the Zeeman energy in a quadrupole
trap is $6k_{\rm B}T$.} \label{fig:sympa-big}
\end{figure}

The MQDT approach makes it feasible to use a larger basis than was
possible in ref.\ \cite{Wallis:LiNH:2011} and to carry out the
calculations on a much finer grid of energies and fields. Figure
\ref{fig:sympa-big} shows the results obtained from MQDT with
perturbative long-range corrections for a converged
basis set with $n_{\rm max}=10$
and $L_{\rm max}=8$, with cross sections calculated on a $51
\times 51$ grid of energies and fields. The resulting coupled-channels
basis sets contains 1887 basis functions, as
compared to 937 functions for the smaller basis set used in Figure
\ref{fig:sympa}, and each coupled-channel calculation is therefore a
factor of 8 more expensive.

As described in ref.\ \cite{Wallis:LiNH:2011}, the elastic and
inelastic cross sections are a strong function of both potential
scaling and basis set size, because they depend sensitively on the
positions of near-dissociation levels. Because of this, calculations on
a single potential do not give quantitative predictions for the ratio
of elastic and inelastic cross sections, and it is essential to explore
the potential-dependence of the results. We found that using the
unscaled potential with a converged basis set gave a highly
atypical contour plot, because it has an accidentally near-zero
scattering length and therefore a very small elastic cross section. The
calculations in Figure \ref{fig:sympa-big} used a potential with an
overall scaling factor of 0.995, which produces a much more typical
contour plot. Exploring the dependence of the results on the scaling
factor confirmed the conclusions of ref.\ \cite{Wallis:LiNH:2011}, that
sympathetic cooling of NH by Li is likely to succeed if the molecules
can be precooled to a temperature around 20~mK.

\section{Conclusions}

We have demonstrated that Multichannel Quantum Defect Theory (MQDT) can provide
quantitatively accurate cross sections for cold and ultracold elastic and
inelastic collisions in magnetic fields for a strongly coupled molecular
system, Li + NH. However, the choice of the phases of the MQDT reference
functions is crucial. For Mg+NH, it was sufficient to choose the phases so that
$Y_{ii}=0$ in every channel included in MQDT \cite{Croft:MQDT2:2012}. For
Li+NH, however, this does not guarantee that all closed-channel poles are moved
far away in energy, and the poles can cause problems in interpolation. In the
present paper, we have developed an improved approach for optimizing the phases
that ensures that closed-channel poles are far away from a reference energy.

We have been able to reproduce the results of coupled-channel calculations
across the entire range of energies and field relevant to sympathetic cooling
of NH by Li, using our new version of MQDT combined with perturbative
corrections for long-range inelasticity caused by the magnetic spin dipolar
interaction. The MQDT results required coupled-channels calculations at only 5
combinations of energy and field, whereas the coupled-channels calculations
\cite{Wallis:LiNH:2011} required 204 combinations. MQDT thus has enormous
potential as an efficient computational method for molecular collisions.

\section{Acknowledgments}

We are grateful to Paul Julienne for many valuable discussions. JFEC is
grateful to EPSRC for a High-End Computing Studentship. The authors are
grateful for support from EPSRC and from EOARD Grant FA8655-10-1-3033.

\bibliography{james,james2,../all}
\end{document}